\documentclass[
    ,final            
  ]
  {aipproc}

\layoutstyle{6x9}

\begin{document}

\title{Discovery of the magnetic field of the B1/B2V star $\sigma$ Lupi}


\author{H.F. Henrichs}{address={
University of Amsterdam, Amsterdam, Netherlands}
}

\author{K. Kolenberg}{
  address={Universit\"{a}t Wien, Vienna, Austria}
}

\author{B. Plaggenborg}{address={
University of Amsterdam, Amsterdam, Netherlands}
}

\author{S.C. Marsden}{address={
Anglo-Australian Observatory, Australia}
}

\author{I.A.~Waite}{address={
University of Southern Queensland, Toowoomba, Australia}
}

\author{J. Landstreet}{address={
University of Western Ontario, London, Canada}
}

\author{J. Grunhut}{address={
Royal Military College of Canada, Kingston, Ontario, Canada}
}

\author{M. Oksala}{address={
University of Delaware, Newark, DE, USA}
}

\author{G. Wade}{address={
Royal Military College of Canada, Kingston, Ontario, Canada}
}

\author{the~MiMeS Collaboration}{address={
http://www.physics.queensu.ca/$\sim$wade/mimes/MiMeS$\_ \_$Magnetism$\_$in$\_$Massive$\_$Stars.html}
}

\begin{abstract}
In our search for new magnetic massive stars we use the strongest indirect indicator of a magnetic field in B stars, which is periodic variability of UV stellar wind lines occurring in a velocity range symmetric around zero.
Our aim is to obtain follow-up spectropolarimetry to search for a magnetic field in magnetic candidate stars.
We quantify UV wind line variability, and analyse its time behaviour. The B1/B2V star $\sigma$ Lupi emerged as a new magnetic candidate star. AAT spectropolarimetric measurements with SEMPOL were obtained.

The stellar wind line variations of $\sigma$ Lupi are similar to what is known in magnetic B stars, but no periodicity could be determined. We detected a longitudinal magnetic field with varying strength and amplitude of about 100 G with error bars of typically 20 G, which supports an oblique magnetic-rotator configuration. The equivalent width variations of the UV lines, the magnetic and the optical line variations are consistent with the well-known photometric period of 3.02 days, which we identify with the rotation period of the star.  Additional observations with ESPaDOnS attached to the CFHT strongly confirmed this discovery, and allowed to determine a precise magnetic period. Further analysis revealed that $\sigma$ Lupi is a helium-strong star, with an enhanced nitrogen abundance and an underabundance of carbon, and has a spotted surface.

We conclude that $\sigma$ Lupi is a magnetic oblique rotator, and is a He-strong star.  It is the fourth B star for which a magnetic field is discovered from studying only its wind variability. Like in the other magnetic B stars the wind emission originates in the magnetic equator, with maximum emission occurring when a magnetic pole points towards the Earth. The 3.01819 d magnetic rotation period is consistent with the photometric period, with maximum light corresponding to maximum magnetic field.
A full paper will be submitted to {\it Astronomy \& Astrophysics}. 

\end{abstract}

\maketitle


\section{Introduction}
High-resolution spectropolarimeters covering a wide wavelength range (Musicos, SEMPOL, ESPaDOnS, Narval, Sophin) allow to detect organized magnetic fields in B and O stars.
The MiMes Collaboration has as primary goal to search systematically for these fields.
It appears that in nearly all magnetic OB stars the dipole component is dominant. In general, these objects act as oblique rotators. The outflowing stellar wind is perturbed by the surface magnetic field, and is periodically modified. In fact, the discovery of a number of magnetic early-type stars was preceded by the discovery of strictly periodic wind variability as observed in the UV, which appeared to be the strongest indirect indicator for the presence of a magnetic field. By this method three magnetic B stars have been found: $\beta$ Cep, $\zeta$ Cas and V2052 Oph, with rotation periods of 12 d, 5.4 d and 3.6 d, respectively.
We describe here the discovery of the magnetic field, the period analysis and abundance determination of $\sigma$ Lup.

The adopted stellar parameters (partly based on \cite{levenhagen:2006, jerzykiewicz:1992}) are:
Spectral Type B1/B2V, $V = 4.416$, $d = 176 ^{+26}_{-20}$ pc, log($L$/$L_{\odot})=3.76 \pm 0.06$, $T_{\rm eff} = 23 000 \pm 550 $ K, log$g = 4.02 \pm 0.10 $ cms$^{-2}$, $R/R_{\odot} = 4.8 \pm  0.5$, $M/M_{\odot} = 9.0 \pm 0.5 $, log Age(y) $= 7.13 \pm 0.13$, $v$sin$i= 68 \pm 6 $ km\,s$^{-1}$,
$P_{\rm phot} =3.0194 \pm 0.0002 $ d, $P_{\rm magn} =3.01819 \pm 0.00034 $ d, and $v_{\rm rad} = 0.0 \pm 0.5 $km\,s$^{-1}$.

\section{Stellar Wind Variability}

We examineded the 4 usable archival IUE spectra, taken between 1992 and 1993, near the Si IV, C IV and N V lines, the prime  stellar wind indicators. Fig.\ 2a shows an overplot along with the significance of the variations. Significant variability was found, similar to what exclusively is found in magnetic stars. For EW variations see Fig.\ 2b (top).

The three previously discovered magnetic B stars, mentioned above, showed a double sine wave in the equivalent width of the UV wind lines, with maximum emission (minimum EW) coinciding with maximum field strength, i.e.\ when a magnetic pole is pointing towards the observer. The expected curve (with arbitrary scaling, see Fig.\ 2b, top) suggests similar behavior for $\sigma$ Lup as well. This supports a model with the emitting material in the magnetic equator.

\section{Spectropolarimetry and magnetic results}

\begin{figure}[h]
$
\begin{array}{cc}
\includegraphics[height=.34\textwidth]{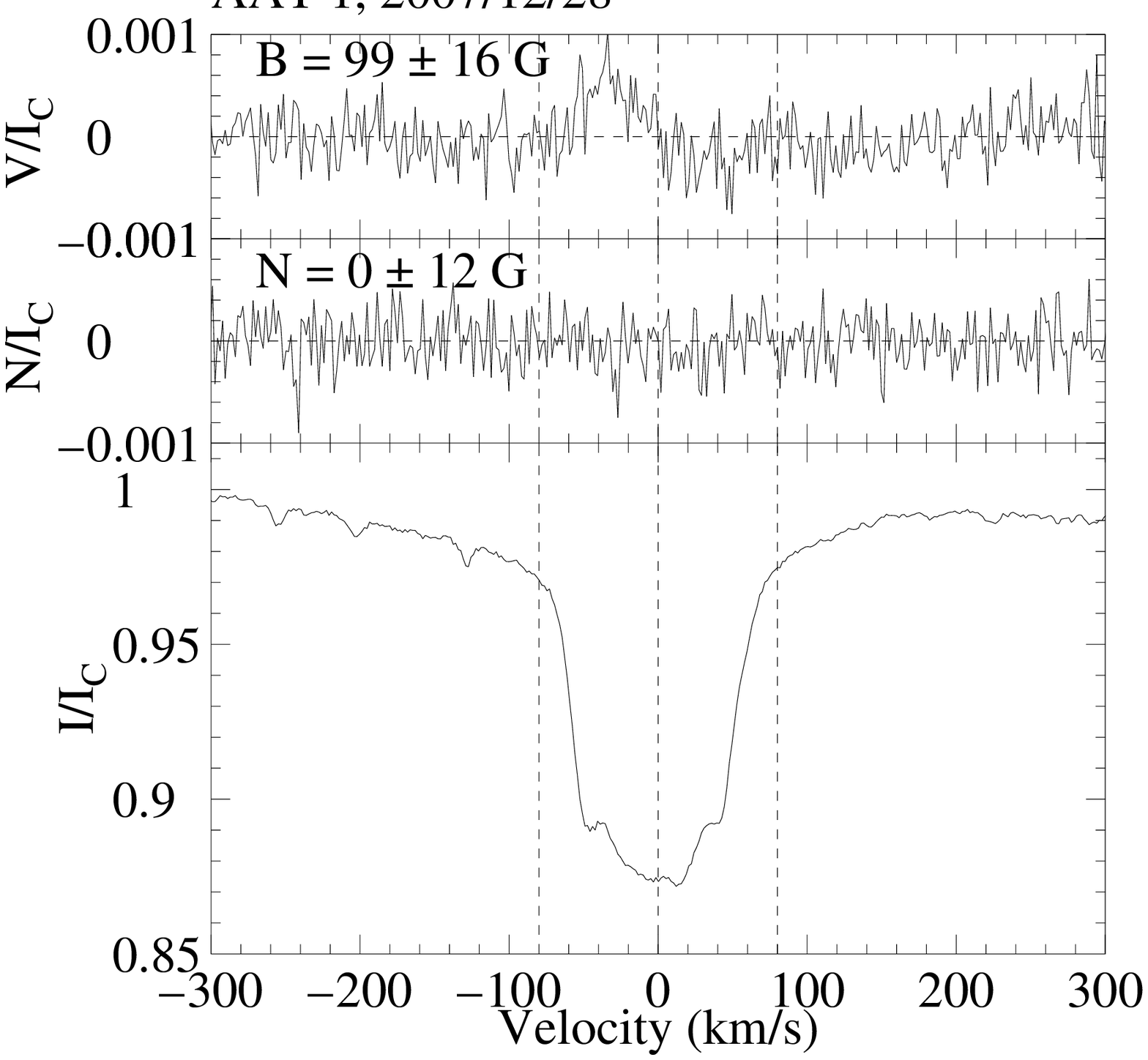} &
\includegraphics[height=.34\textwidth]{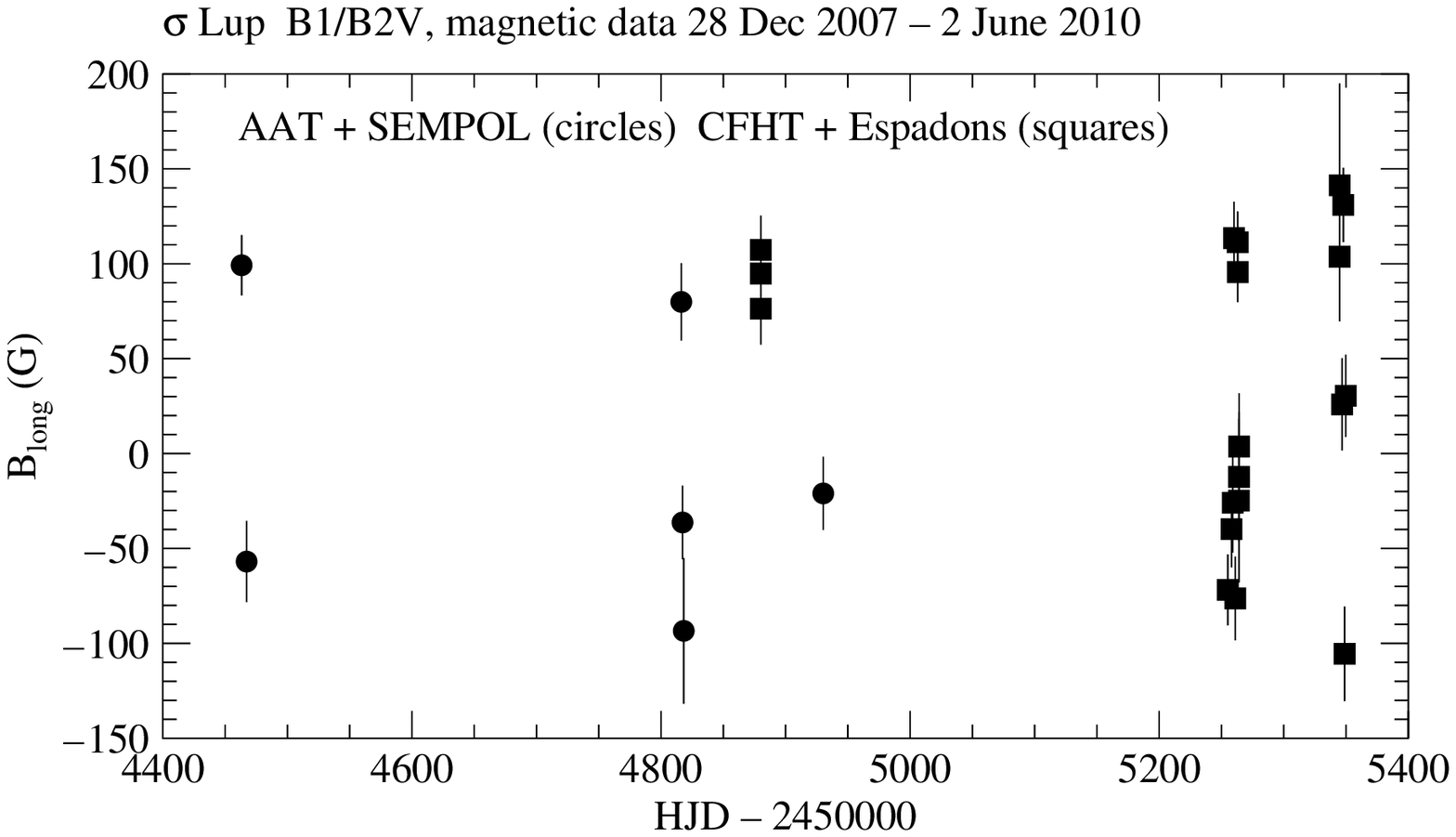}
\end{array}$

\caption{(a) The discovery magnetic measurement of $\sigma$ Lupi at AAT. Displayed are LSD Stokes unpolarised $I$, (null) $N$ profile (for integrity purposes) and circularly polarised $V$ profiles of SEMPOL spectra. The integrated Zeeman signature in the $V$ profile over the width of the line (within the outer dashed vertical lines) gives the value of the longitudinal component of the magnetic field, integrated over the stellar surface: $B_l= 99 \pm16$ G. (b) Magnetic data of $\sigma$ Lup.}
\end{figure}

The échelle spectropolarimeters SEMPOL attached to the 3.9 m AAT in Australia and ESPaDOnS at the 3.6m CFHT at Hawaii were used. The reduction was provided by the proprietary software of the observatory, called Libre ESpRIT version 2.06, see \cite{donati:2006}.
Weak stellar magnetic fields can be detected through the Zeeman signatures generated in the shape and polarisation state of spectral line profiles, applying a cross-correlation technique the Least-Squares Deconvolution (LSD). The LSD method combines the very small circularly polarised signatures, properly weighted, of all available line profiles in the spectrum to increase the signal to noise ratio. Typical exposure times are $4 \times 200$ to $4 \times 400$ sec. For $\sigma$ Lup we used 171 spectral lines to obtain a mean Stokes $V$ profile. When a magnetic field is present, the Stokes $V$ profile indicates a Zeeman signature. See Fig.\ 1a for the discovery Stokes $V$ spectra.

\begin{figure}[h]
$
\begin{array}{cc}
\includegraphics[height=.49\textheight]{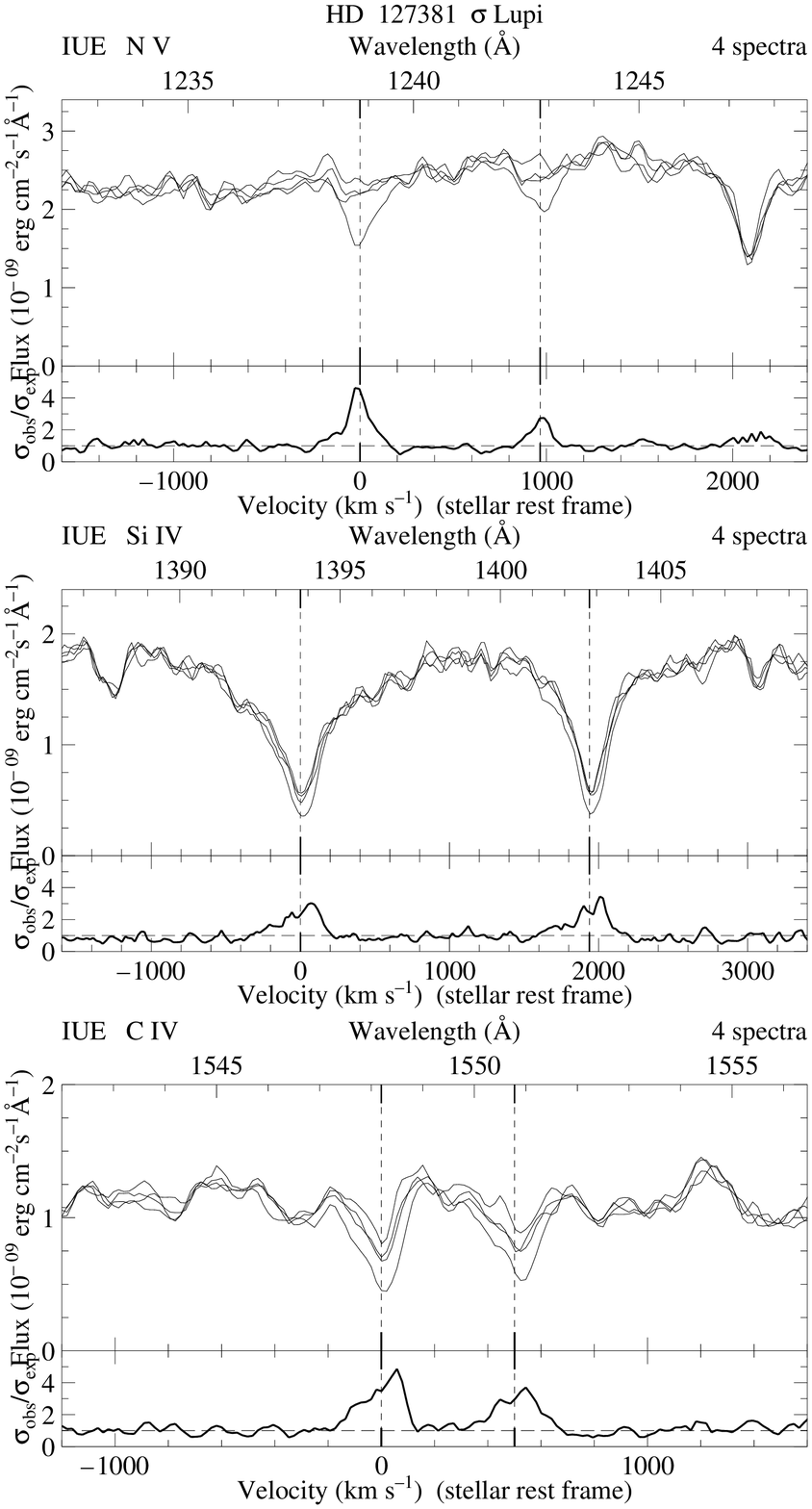} &
\includegraphics[height=.49\textheight]{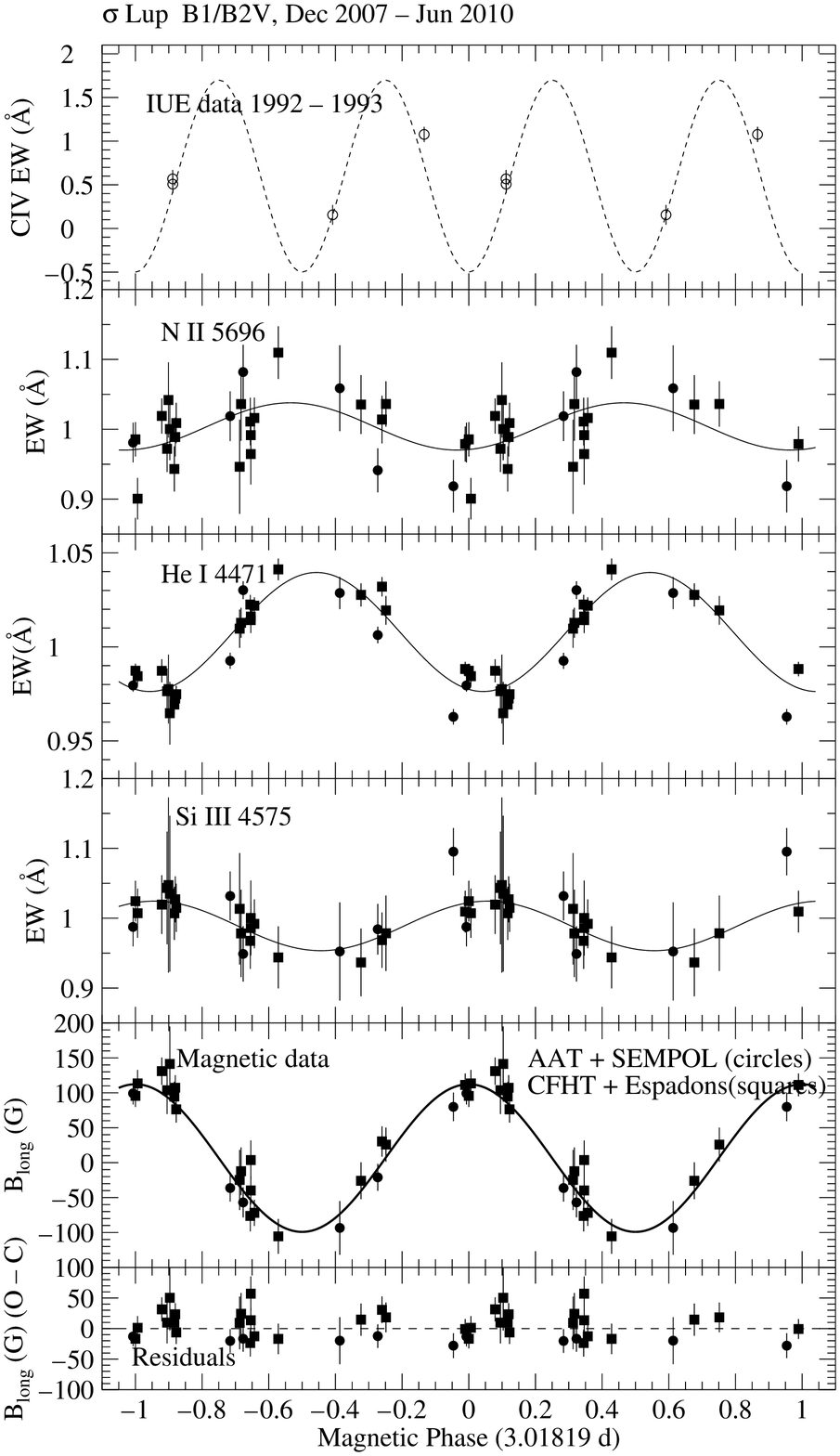}
\end{array}$

\caption{(a) Overplot of variable UV wind lines of $\sigma$ Lupi. Such variability is exclusively found in magnetic B stars, which prompted a magnetic search. (b) Overplot of all magnetic and EW data folded with the adopted period of 3.0186 d. Lower two panels: magnetic data with their residuals with  best-fit cosine curve. Middle panels EW of the Si III 4575, He I 4471 and N II 5696 lines with a best-fit sinusoid superposed. Note
the opposite behavior of the different lines. Top: EW the CIV 1540 doublet 15 years earlier (Fig. 2a) with a suggested double sine wave, similar to what is observed in other magnetic B stars.}
\end{figure}

From the LSD spectra we computed the mean longitudinal field ($B_l$), integrated over the stellar surface. The results are plotted in Fig.\ 1b. The smallest error bars are about 16 G.
A best fit with reduced $\chi^2= 1.0$ of the function
$B_l(t) = B_0 + B_{\rm max} \cos(2\pi(t-t_0)/P)$ gives:
$B_0= (6 \pm 5)$ G, $B_{\rm max}= (106 \pm 7)$ G, $P= 3.01858 \pm 0.00015$ d
and $t_0= $JD$ 2455103.12 \pm 0.567$.
This function is overplotted in Fig.\ 2b. 

With the stellar parameters above it follows that $i> 50^\circ$. The magnetic tilt angle $\beta$ is then constrained by the observed ratio $B_{\rm max} /B_{\rm min} = \cos(i +\beta)/\cos(i -\beta) = -1.15^{+0.23}_{-0.29}$, which implies that $\beta$ is close to $90^\circ$. This gives a polar field of $\sim 400$ G.

\section{Period and Abundance Analysis}
The photometric period is $3.0186\pm0.0004$ d, which was determined 3147 cycles earlier.
The extrapolated epoch of maximum light coincides within the uncertainties with the epoch of maximum (positive) magnetic field.
Our new photometry attempts to confirm this result was unsuccessful because of the lack of suitable nearby reference stars.

EW variations (Fig.\ 2b) in phase with the magnetic field is found in all studied Si lines. The He lines vary strongly in antiphase. Abundances were determined at rotational phase 0.42, when the He lines are strongest (see Fig.\ 3).
The star appears to be a He-strong star, with overabundance of N and underabundance of C, similar to what is found in most other magnetic early-type B stars.
\begin{figure}[h]
\includegraphics[height=.37\textwidth]{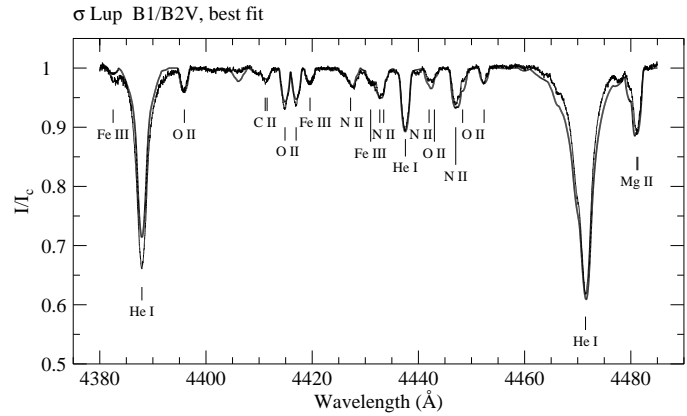}
\caption{Portion of the optical spectrum of $\sigma$ Lup taken at phase 0.418, (with strongest He lines) with most lines identified, overplotted with the modelfit (thick line) to determine abundances.}
\end{figure}
\section{Conclusions}
$\sigma$ Lup is the fourth magnetic B star found by its stellar wind variability.
The polar field is $\sim 400$ G. With the 3.0 d period it is a relatively fast magnetic rotator.
The star is a He-rich star, with N-enhancement.
The photometric period being the same as the magnetic period could serve as another indirect indicator for hosting a magnetic field.
The EW variations of the optical lines indicate significant spots on the stellar surface.



\bibliographystyle{aipproc}   

\bibliography{references}

\IfFileExists{\jobname.bbl}{}
 {\typeout{}
  \typeout{******************************************}
  \typeout{** Please run "bibtex \jobname" to optain}
  \typeout{** the bibliography and then re-run LaTeX}
  \typeout{** twice to fix the references!}
  \typeout{******************************************}
  \typeout{}
 }

\end{document}